# Constrained Trading Networks

Can Kizilkale[1] & Rakesh Vohra[2]

[1]Computational Research Division, Lawrence Berkeley National Laboratory

[2]Department of Economics & Department of Electrical and Systems Engineering,

University of Pennsylvania

June 22, 2020

## Abstract

Trades based on bilateral (indivisible) contracts can be represented by a network. Vertices correspond to agents while arcs represent the non-price elements of a bilateral contract. Given prices for each arc, agents choose the incident arcs that maximize their utility. We enlarge the model to allow for polymatroidal constraints on the set of contracts that may be traded which can be interpreted as modeling limited one-for-one substitution. We show that for two-sided markets there exists a competitive equilibrium however for multi-sided markets this may not be possible.

## 1 Introduction

Trades based on bilateral (indivisible) contracts are represented by a network. The vertices correspond to agents while the arcs represent the non-price elements of a bilateral contract. The arc's orientation identifies which agent is the "buyer" and which the "seller". The model is rich enough to allow an agent to be a buyer in some trades and a seller in others. It subsumes the classic assignment model (Shapley and Shubik 1971). Given prices for each arc/trade, an agent chooses the subset of incident arcs that maximize her utility. A central question is whether a competitive equilibrium in this economy exists.

In general it does not. Assuming quasi-linearity and a *full substitutability* condition on agents' preferences, (John William Hatfield et al. 2013) have shown that a competitive equilibrium exists. Full substitutability is a generalization of the gross-substitutes property introduced by Kelso-Crawford and is equivalent to valuations over contracts being $M^\sharp$ concave (Fujishige and Yang 2003, Murota and Tamura 2003). Competitive equilibria of trading networks are also stable outcomes in that they cannot be blocked by any coalition of agents and trades. A blocking set is a set of contracts and corresponding prices such that all agents party to these contracts (strictly) prefer them (while possibly declining some of their equilibrium contracts)(John W Hatfield et al. 2013). Conversely, in any stable outcome it is possible to set prices for contracts not involved in this outcome, to support the outcome as a competitive equilibrium.

---

[1]Research supported in part by DARPA grant HR001118S0045.

Candogan, Epitropou, and Vohra 2016, show that these results are a consequence of the optimality conditions of a generalized submodular flow problem in a suitable network. The optimal solutions to this flow problem (and its dual) yield a competitive equilibrium outcome and supporting prices.

Contracts in the trading networks model are exogenously given and it is assumed that an agent can participate in *any* subset of contracts to which she is a party to. In other words, a vertex is free to choose any subset of its incident arcs. This is unsatisfying if the contracts under consideration involve the exchange of resources in limited supply. Certain combinations of contracts could be infeasible. This can be modeled by defining an agents value for infeasible subsets of contracts to be $-\infty$, which may result in a violation of $M^\sharp$ concavity.

In this paper we enlarge the trading networks model to allow for explicit constraints on the set of contracts that may be traded. It is obvious that for arbitrary constraints, it is unlikely that a competitive equilibrium will exist.[1] Therefore, we restrict attention to constraints that can be represented by polymatroids. Polymatroids are polyhedrons associated with integer valued submodular functions. They play an important role in combinatorial optimization but also make an appearance in some market design settings. In the context of trading networks they can be interpreted as modeling limited one-for-one substitution. Informally, the marginal rate of technical substitution between two products evaluated at any bundle is either zero or one. This interpretation appears in (Milgrom 2009).

The first contribution of this paper is establishing the existence of a competitive equilibrium for two sided markets, ones in which agents are either sellers or buyers but not both.[2] Closest comparable results we are aware of are (Kojima, Sun, and Yu 2018) and (Gul, Pesendorfer, and Zhang 2019). Both consider one-sided markets, the first without transfers and the second with. Theorem 3 of the second paper is a special case of our result. Our second result shows that existence of a competitive equilibrium fails in multi-sided settings.

The next section introduces the notation used in this paper. The subsequent section summarizes result needed from the literature on discrete convexity. The remaining sections describe our main results.

## 2 Notation

A trading network is represented by a directed multigraph $G = (N, E)$ where $N$ is the set of vertices and $E$ the set of arcs. Each vertex $i \in N$ corresponds to an agent and each arc $e \in E$ corresponds to the non-price elements of a trade that can take place between the incident pair of vertices. For each $e \in E$, the source vertex $e^+$ corresponds to the seller and the sink vertex $e^-$ corresponds to the buyer in the trade. Let $\delta_+(i)$ and $\delta_-(i)$ be the outgoing and incoming arcs incident to vertex $i \in N$, and $\delta(i) = \delta_+(i) \cup \delta_-(i)$. So as to avoid having to distinguishing between a trade involving one unit of a good and two units of the same good, we will allow for integral 'intensities' of trade. Formally, an **outcome** is any vector $x \in \mathbb{Z}_+^E$ where $x_e$ denotes the number of copies of trade $e$ that were executed. We define a price vector $p \in \mathbb{R}^E$, where $p_e$ is the price associated with the trade that corresponds to the arc $e$. Denote by $p^X$ the price vector restricted to the arcs in $X$.

---

[1] This was already observed in (John W Hatfield et al. 2013).

[2] This is equivalent to a multi-sided setting where the utility of the corresponding agents is separable over the contracts it buys and sells though.

It is more convenient to represent the volume of trade associated with arc $e$ between agents $i$ and $j$ using two variables rather than one: $y_e^i$ and $y_e^j$. If $e \in \delta_+(i)$ then, $y_e^i \geq 0$ otherwise $y_e^i \leq 0$. We will call the $y$ variables **net-flows**.

Every $x \in \mathbb{Z}_+^E$ that is an outcome, corresponds to a new flow. For any arc $e$ between agents $i$ and $j$ we have $|y_e^i| = |y_e^j| = x_e$ and $y_e^i + y_e^j = 0$. A net-flow $y = (y^1, \ldots, y^{|N|})$ is called feasible if it corresponds to an outcome, i.e., for every arc $e$ we have $y_e^i + y_e^j = 0$ whenever $e \in \delta_+(i) \cap \delta_-(j)$. If $y^i$ is the incidence vector of trades in which $i$ is a part of, the value function of agent $i$ will be written as $w_i : y^i \to \mathbb{R} \cup \{-\infty\}$.[3] Agent $i$'s surplus or utility for the vector of trades $y^i$ will be

$$u_i(y^i, p) = w_i(y^i) + \sum_{e \in \delta_+(i)} y_e^i p_e + \sum_{e \in \delta_-(i)} y_e^i p_e.$$

Given a price vector $p \in \mathbb{R}^{\delta(i)}$, agent $i \in N$'s demand correspondence is

$$D_i(p) = \arg\max\{u_i(y^i, p) : y^i \in \mathbb{Z}_+^{\delta_+(i)} \times \mathbb{Z}_-^{\delta_-(i)}\}.$$

DEFINITION 2.1 *A feasible net-flow $y = (y^1, \ldots, y^{|N|})$ along with a price vector $p \in \mathbb{R}^E$ is a* **competitive equilibrium** $(y, p)$ *if, for all $i \in N$, $y^i \in D_i(p)$.*

DEFINITION 2.2 *An* **efficient** *outcome is one that solves the following problem:*

$$\max \sum_{i \in N} w_i(y^i)$$

$$\text{s.t. } y_e^i + y_e^j = 0 \ \forall \ i = e^+, \ j = e^- \ \forall e \in E$$
$$y_e^i \in \mathbb{Z}_+ \ \forall e \in \delta_+(i) \ \forall i$$
$$y_e^i \in \mathbb{Z}_- \ \forall e \in \delta_-(i) \ \forall i$$

Under quasi-linearity, every competitive equilibrium outcome $y$ is efficient and conversely. Further, if $y$ is efficient, there exists a price vector $p$ such that the pair $(y, p)$ is a competitive equilibrium.

## 2.1 Background on Discrete Convexity

In this section we review definitions and properties of $M$-convex sets/functions. For a more extensive discussion see (Murota 2003b). For all $y \in Z^n$ we define $supp^+(y) = \{i | y_i > 0\}$ and $supp^-(y) = \{i | y_i < 0\}$. The notion of $M-convexity$ is based on the exchanges between the sets $supp^+$ and $supp^-$. Let's start by defining the "M-convex" set.

---

[3]Having $-\infty$ in the range of the value function allows for incorporating trading constraints as described in John W Hatfield et al. 2013, e.g., if a trader cannot sell goods without procuring its inputs first, this can be incorporated by specifying $-\infty$ for bundles of trades where this happens.

DEFINITION 2.3 *A set $B \subseteq \mathbb{Z}^E$ is called **M-convex** if for all $x, y \in B$ and for all $u \in supp^+(x-y)$ there exists a $v \in supp^-(x-y)$ such that $x - \chi_u + \chi_v$ and $y + \chi_u - \chi_v$ are both in the set $B$.*

In definition 2.3, $\chi_u$ refers to the basis vector where the $u^{th}$ element is one and all others are zero. M-convex functions are defined similarly in (2.4).

DEFINITION 2.4 *A function $f : \mathbb{Z}^E \to R$ for an M-convex domain $B$ is called **M-convex** if for all $x, y \in B$, $\forall u \in supp^+(x-y)$ there exists a $v \in supp^-(x-y)$ such that*

$$f(x) + f(y) \geq f(x - \chi_u + \chi_v) + f(y + \chi_u - \chi_v).$$

DEFINITION 2.5 *A function $f$, is called **M-concave** if $-f$ is M-convex.*

M-convexity has strong ties with submodularity. A real valued submodular function $f$ defined on $E$ is one that satisfies $f(S) + f(T) \geq f(S \cup T) + f(S \cap T)$ for all $S, T \subseteq E$.

THEOREM 2.1 *For any M-convex set $B$, the function $f(S) = \max\{\sum_{j \in S} x_j | x \in B\}$ for every $S \subseteq E$ is a submodular function.*

Theorem 2.1 gives a connection between M-convex sets and integral submodular set functions. The converse is also true.

COROLLARY 2.1 *If $f$ is an integer valued submodular function on $E$, then there exists an M-convex set $B \subseteq \mathbb{Z}^E$ such that $f(S) = \max\{\sum_{j \in S} x_j | x \in B\}$.*

If $f$ is a submodular function on $E$, then, the polymatroid associated with $f$ is the polyhedron $P_f = \{x \in \mathbb{R}^E : \sum_{e \in S} x_e \leq f(S) \; \forall S \subseteq E\}$. It is called an integral polymatroid if $f$ is integer valued. In fact, the M-convex set $B$ in Corollary 2.1 is precisely $P_f$.

The definition of M-convexity relies on the exchange between two different elements and it can be weakened to cover single element exchanges. We start with the set definition first.

DEFINITION 2.6 *A set $B \subseteq \mathbb{Z}^E$ is called $M^\sharp$-**convex** if for all $x, y \in B$ and for all $u \in supp^+(x-y)$ we have one of the following:*

- *$x - \chi_u$ and $y + \chi_u$ are both in $B$.*

- *There exists a $v \in supp^-(x-y)$ such that $x - \chi_u + \chi_v$ and $y + \chi_u - \chi_v$ are both in the set $B$.*

$M^\sharp - convex$ functions are defined as follows.

DEFINITION 2.7 *A function $f : \mathbb{Z}^E \to R$ is called $M^\sharp$-**convex** if for all $x, y \in B$, $\forall u \in supp^+(x-y)$ we have one of the following.*

- *$f(x) + f(y) \geq f(x - \chi_u) + f(y + \chi_u)$.*

- *There exists a $v \in supp^-(x-y)$ such that $f(x) + f(y) \geq f(x - \chi_u + \chi_v) + f(y + \chi_u - \chi_v)$.*

A function $f : \mathbb{Z}^E \to R$ is called $M^\sharp$-**concave** if $-f$ is $M^\sharp$-convex. The **domain** of a function $f : \mathbb{Z}^E \to R$ is defined to be $dom f = \{x \in Z^n | f(x) < \infty\}$. A consequence of a function being $M^\sharp$-convex is the following:

LEMMA 2.1 *The domain of an $M^\sharp$-convex function is an $M^\sharp$-convex set.*

DEFINITION 2.8 *The **convex-extension** of a function is as follows (the intersection of the supporting hyperplanes of the epigraph):*

$$\hat{f}(x) = \min\{\sum_z \lambda_z f(z) | \sum_z \lambda_z z = x, 0 \leq \lambda_z \leq 1, \sum_z \lambda_z = 1\}$$

We can define **concave-extension** similarly.

DEFINITION 2.9 $\hat{f}(x)$ *is the **concave-extension** of $f(x)$ if $-\hat{f}(x)$ is the convex-extension of $-f(x)$.*

## 2.2 Optimality

Suppose each agent's valuation function, $w_i$ is $M^\sharp$-concave and let $f(y) = \sum_i w_i(y^i)$. Note, $f(y)$ is also $M^\sharp$ concave because the argument of each $w_i$ are disjoint. The problem of finding an efficient outcome in the trading network setting can be formulated as:

$$\text{maximize} : f(y) \text{ s.t. } y \in B.$$

Here $B$ represents the set of feasible trades:

$$y_e^i + y_e^j = 0 \; \forall \; i = e^+, \; j = e^- \; \forall e \in E$$

$$y_e^i \in \mathbb{Z}_+ \; \forall e \in \delta_+(i) \; \forall i$$

$$y_e^i \in \mathbb{Z}_- \; \forall e \in \delta_-(i) \; \forall i$$

A convex relaxation of this problem is:

$$\text{maximize} : \hat{f}(y) \text{ s.t. } y \in B^* \tag{1}$$

where $\hat{f}(y)$ is the concave extension (definition 2.8) of the objective function $f(y)$ and $B^*$ is the set of solutions to

$$y_e^i + y_e^j = 0 \; \forall \; i = e^+, \; j = e^- \; \forall e \in E$$

$$y_e^i \geq 0 \; \forall e \in \delta_+(i) \; \forall i$$

$$y_e^i \leq 0 \; \forall e \in \delta_-(i) \; \forall i$$

It is known that problem (1) has an integral solution (Murota 2003a), the Lagrange multipliers associated with the inequality representation of $B$ will be supporting prices and this establishes the existence of a competitive equilibrium.

The set $B^*$ in problem (1) had a particular structure one might wonder if integrality of the convex relaxation might hold more generally. Here we show that it does if the set of net-flows is constrained to lie in an integral polymatroid, denoted $P$. Consider:

$$\text{maximize } \hat{f}(y) \text{ s.t. } y \in P \tag{2}$$

We show problem 2 has an integer optimal. This result, is, as far as we know new, however special cases of it do appear, for example (Gul, Pesendorfer, and Zhang 2019).

Let us recall the following theorem.

THEOREM 2.2 *(Thm 4.22 Murota 2003a) For M-convex sets $B_1$ and $B_2$ the convex closure of their intersection is equal to the intersection of their convex closures, that is*

$$\bar{B}_1 \cap \bar{B}_2 = \overline{(B_1 \cup B_2)}.$$

DEFINITION 2.10

$$B_f(x) = \{z \in N(x) \| \exists \lambda, \sum_z \lambda_z = 1, \sum_z \lambda_z z = x, \sum_z \lambda_z f(z) = \hat{f}(x)\}$$

where $N(x)$ is the unit hypercube containing $x$ and $\hat{f}(x)$ is the convex extension of function $f$.

$B_f(x)$ will have an important role in our analysis. The convex extension of function $f(y)$ is a piece-wise linear function and $B_f(x)$ represent the maximal set that contains $x$ such that convex extension is linear. This can be viewed as the facets of the epigraph.

LEMMA 2.2 *For every $M^\sharp$ concave function $f$ and $x$ in the convex hull of the domain of $f$, $B_f(x)$ is an M-convex set.*

**Proof.** Let $x_1, x_2 \in B_f(x)$. Since $f(x)$ is an $M^\sharp$ concave function there exists $u \in supp_+(x_1 - x_2)$ and $v \in supp_-(x_1 - x_2)$ such that,

$$f(x_1 - \chi_u + \chi_v) + f(x_2 + \chi_u - \chi_v) \geq f(x_1) + f(x_2).$$

Now take convex coefficients $\lambda^1, \lambda^2$ such that

$$\sum_{z \in N(x)} \lambda_z^1 z = \sum_{z \in N(x)} \lambda_z^2 z = x$$

and $\sum_{z \in N(x)} \lambda^i f(z) = \hat{f}(x)$ (by definition such a $\lambda^i$ exists). Then for $\lambda = \frac{\lambda^1 + \lambda^2}{2}$ we have $\sum_z \lambda_z z = x$, $\sum_{z \in N(x)} f(z) = \hat{f}(x)$, $\lambda_{x_1} > 0$ and $\lambda_{x_2} > 0$. If we set $\lambda_\delta = \min\{\lambda_{x_1}, \lambda_{x_2}\}$ we obtain the following inequality

$$\hat{f}(x) \geq [\sum_{z \in N(x) - \{x_1, x_2\}} \lambda_z f(z)] + \lambda_\delta(f(x_1 - \chi_u + \chi_v) + f(x_2 + \chi_u - \chi_v)) + (\lambda_{x_1} - \lambda_\delta)f(x_1) + (\lambda_{x_2} - \lambda_\delta)f(x_2)$$

$$\geq \sum_{z \in N(x)} f(z) = \hat{f}(x).$$

But then, by definition $(x_1 - \chi_u + \chi_v), (x_2 + \chi_u - \chi_v)$ are both in $B_f(x)$ hence $B_f(x)$ is an M-convex set. ∎

COROLLARY 2.2 *For all probability vectors $\lambda$ such that $\sum_{z \in B_f(x)} \lambda_z z = x$ we have $\hat{f}(x) = \sum_z \lambda_z f(z)$.*

Proof follows trivially from the definition of $B_f(x)$.

THEOREM 2.3 *If $f$ is $M^\#$-concave function and $P$ is an integral polymatroid, then, there exists an integer optimizer for problem 2.*

**Proof.** Let $x$ be an optimizer for problem 2. If $N(x) \subseteq B$ then we are done since at least one extreme point of $N(x)$ will also be an optimizer. Else let $F_x$ be the face of $B$ that contains $x$. $F_x$ is an M-convex set (as we have seen earlier M-convex sets are essentially polymatroids). From lemma 2.2 we know that $B_f(x)$ is an M-convex set and from theorem 2.2 we have

$$x \in \overline{F_x} \cap \overline{B_f(x)} = \overline{(F_x \cup B_f(x))}.$$

But then there exists convex coefficients $\lambda$ such that

$$\sum_{z \in (F_x \cup B_f(x))} \lambda_z z = x.$$

From corollary 2.2 we have

$$\sum_{z \in (F_x \cup B_f(x))} \lambda_z f(z) = \hat{f}(x).$$

But then there exists a $y^* \in (F_x \cup B_f(x))$ such that $f(y^*) \leq \hat{f}(x)$ and since $x$ is an optimizer, $y^*$ has to be an optimizer too. Since $y^*$ is integral this concludes the proof. ∎

## 3 Separable Utility

Consider the case of one buyer (agent 1) and one seller (agent 2). Thus, the vertex corresponding to agent 1 has no outgoing arcs only incoming ones. The vertex corresponding to agent 2 has no incoming arcs only outgoing arcs. Hence, $y^1 = -y^2 = x$ where $x$ is an outcome. Assume that $x$ is constrained to a polymatroid $P$. $w_1(y^1)$ is $M^\#$-concave but $w_2(y^2)$ is linear. Denote by the $\hat{w}_i$ the convex extension of $w_i$. Our objective is to maximize

$$\hat{w}_1(y^1) + \hat{w}_2(y^2) = \hat{w}_1(x) + \hat{w}_2(-x)$$

over $x \in P$ where $\hat{w}_i$ is the concave extension of $w_i$.

LEMMA 3.1 *The maximizer(s) of*

$$\hat{w}_1(y^1) + \hat{w}_2(y^2) = \hat{w}_1(x) + \hat{w}_2(-x)$$

*over $x \in P$ are integral.*

**Proof.** Let $x$ be a maximizer of the objective. Observe that $w_1$ is linear over the set $B_{w_1}(x)$. Then $w_1(x) + w_2(-x)$ is linear over the set $B_{w_1} \cap P$ (both $w_1$ and $w_2$ are linear on this set). But then, this is a linear program and there exists an extreme point of $B_{w_1} \cap P$ that achieves the optimum, since the intersection of two integral M-convex sets is an integral polyhedra all of whose extreme points are integral. The proof follows. ∎

We generalize Lemma 3.1 to trade networks where each agents utility $w_i(y^i)$ is separable as follows.

$$w_i(y^i) = w_i^+(y^{i+}) + w_i^-(y^{i-})$$

where $w_i^+$ is $M^\sharp$-concave, $w_i^-$ is linear and $y^{i+}, y^{i-}$ represent the net-flows that correspond to the outgoing/incoming edges respectively. The problem of finding an efficient allocation, denoted problem$EA$, can be written as follows.

$$\max \sum_i (w_i^+(y^{i+}) + w_i^-(y^{i-})) \tag{3}$$

$$\text{s.t. } x \in P \tag{4}$$

$$y_e^i = x_e = -y_e^j, \ \forall e \in \delta_+(i) \cap \delta_-(j) \tag{5}$$

where $P$ is an integral polymatroid.

THEOREM 3.1 *Problem EA has an integral optimum.*

**Proof.** For $i \neq j$ we have $\delta_+(i) \cap \delta_+(j) = \emptyset$ and $supp(y^{i-}) \cap supp(y^{j-}) = \emptyset$. Then, $\sum_i (w_i^+(y(\delta_+(i)))$ (represented by $w^+(x)$) is an $M^\#$-concave function while $\sum_i w_i^-(y(\delta_-(i)))$ (represented by $w^-(-x)$) is linear. Substituting $w^+$ and $w^-$ into the objective function allows us to invoke Lemma (3.1) to conclude the proof. ∎

The separability requirement clearly holds in two sided markets where an agent is either a buyer or a seller but not both. Hence, buyers can have $M^\sharp$-utilities over the edges they 'consume' while sellers have linear costs over the edges they 'sell'. The separability requirement can be relaxed. For all $i$ let the utility function be $w_i(y^i) = w_i(y^{i+}, y^{i-})$ such that when $y^{i+}$ is kept constant the function is linear over $y^{i-}$ and when $y^{i-}$ is kept constant it is $M^\sharp$-concave over $y^{i+}$. In other words, when the incoming/outgoing links are kept constant, the function is $M^\sharp$-concave/linear on the outgoing/incoming edges.

# 4 Impossibility

In this section we show that if we relax separability and allow the market to be multi-sided, then, a competitive equilibrium need not exist.

Let the outcome vector $x \in \mathbb{Z}^E$ is constrained to lie in a polymatroid. This means that trade between two distinct pairs of agents can be linked to each other. For example, the volume of trade that a pair $a, b$ can conduct could be limited by the volume of trade that another pair, $c, d$, conduct. This can happen if the execution of trades require the use of shared infrastructure. For instance, $a$ could be selling electricity to $b$ and similarly $c$ is selling electricity to $d$. If they use a common grid electricity to distribute power, then, the trades of one pair constrain the trades that the other pair can engage in.

We append to problem (1) the requirement that the outcome vector, $x \in \mathbb{Z}^E$, lie in a polymatroid. arbitrary polymatroidal constraints on the links ignoring their orientation. We examine whether this augmented problem has an integral solution. The example below shows this is not the case and demonstrates that a competitive equilibrium in this setting does not exist.

EXAMPLE 1 *Consider the network given in figure 1 with two agents, $N_1, N_2$. Each agent has one incoming and one outgoing link denoted $e$ and $g$. The value functions of each agent are displayed below. They are $M^\sharp$-concave.*

$$
\begin{align}
w_1(y_e^1, y_g^1) = & \tag{6} \\
& -0.5 & y_e^1 = 0, y_g^1 = 0 \tag{7} \\
& -0 & y_e^1 = 1, y_g^1 = -1 \tag{8} \\
& -1 & y_e^1 = 1, y_g^1 = 0 \tag{9} \\
& -1 & y_e^1 = 0, y_g^1 = -1 \tag{10} \\
& -\infty & elsewhere. \tag{11}
\end{align}
$$

$$
\begin{align}
w_2(y_e^2, y_g^2) = & \tag{12} \\
& -0.5 & y_e^2 = 0, y_4 = 0 \tag{13} \\
& -0 & y_g^2 = 1, y_e^2 = -1 \tag{14} \\
& -1 & y_g^2 = 1, y_e^2 = 0 \tag{15} \\
& -1 & y_g^2 = 0, y_e^2 = -1 \tag{16} \\
& -\infty & elsewhere. \tag{17}
\end{align}
$$

*Since the arguments of the function $w_1$ and $w_2$ are disjoint, the function $f(y) = w_1(y_e^1, y_g^1) + w_2(y_e^2, y_g^2)$ is also $M^\sharp$-Concave. Consider the polymatroid on $x \in \mathbb{Z}^E$ defined below:*

$$P = \{x \| x_e \leq 1, x_g \leq 1, x_e + x_g \leq 1\}.$$

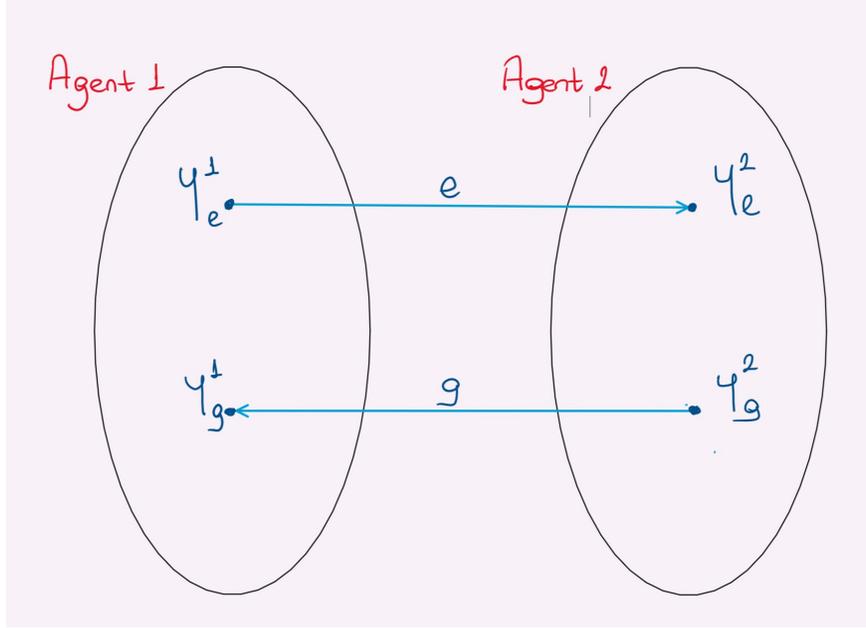

Figure 1: Simple Counterexample

*The net-flows corresponding to $x$ are $y_e^1 = -y_e^2 = x_e$ and $-y_g^1 = y_g^2 = x_g$. Given an outcome $x$, denote by $y(x)$ the corresponding net-flows. Then, we have $f(y(0,0)) = -1, f(y(1,1)) = 0, f(y(1,0)) = f(y(0,1)) = -2$. However,*

$$\max_{x \in P} \hat{f}(y(x)) \geq 1/2 f(y(0,0)) + 1/2 f(y(1,1)) = -0.5$$

*and this is larger than each integral feasible flow which will lead to a larger objective value. This shows that we cannot have an integral efficient outcome.*

## 5  Conclusion

In trading networks even relatively simple constraints on the set of feasible trades are an obstacle to the existence of a competitive equilibrium. Under polymatroidal constraints separability of the utility function is crucial for the existence of competitive equilibrium.